\documentclass[sn-mathphys-num]{sn-jnl}

\usepackage{graphicx}%
\usepackage{multirow}%
\usepackage{amsmath,amssymb,amsfonts}%
\usepackage{amsthm}%
\usepackage{mathrsfs}%
\usepackage[title]{appendix}%
\usepackage{xcolor}%
\usepackage{textcomp}%
\usepackage{manyfoot}%
\usepackage{booktabs}%
\usepackage{algorithm}%
\usepackage{algorithmicx}%
\usepackage{algpseudocode}%
\usepackage{listings}%
\usepackage[normalem]{ulem}

\begin{document}

\title[Article Title]{Observation of Kardar-Parisi-Zhang universal scaling in two  dimensions } 

\author[1]{\fnm{Simon} \sur{Widmann}\textsuperscript{\dag,}}
\author*[1]{\fnm{Siddhartha} \sur{Dam}\textsuperscript{\dag,}}\email{siddhartha.dam@uni-wuerzburg.de}
\author[1]{\fnm{Johannes} \sur{Düreth}}
\author[1]{\fnm{Christian G.} \sur{Mayer}}
\author[2]{\fnm{Romain} \sur{Daviet}}
\author[2]{\fnm{Carl Philipp} \sur{Zelle}}
\author[1]{\fnm{David} \sur{Laibacher}}
\author[1]{\fnm{Monika} \sur{Emmerling}}
\author[1]{\fnm{Martin} \sur{Kamp}}
\author[2]{\fnm{Sebastian} \sur{Diehl}}

\author[1]{\fnm{Simon} \sur{Betzold}}
\author[1]{\fnm{Sebastian} \sur{Klembt}}
\author*[1]{\fnm{Sven} \sur{Höfling}}\email{sven.hoefling@uni-wuerzburg.de}

\affil[1]{\orgdiv{Technische Physik and Würzburg-Dresden Cluster of Excellence ct.qmat}, \orgname{Universität Würzburg}, \orgaddress{\street{Am Hubland}, \city{Würzburg}, \postcode{97074}, \country{Germany}}}

\affil[2]{\orgdiv{Institute for Theoretical Physics}, \orgname{University of Cologne}, \orgaddress{\street{Zülpicher Strasse 77}, \city{Cologne}, \postcode{50937}, \country{Germany}}}

\def\thefootnote{\dag}\footnotetext{These authors contributed equally to this work.}\def\thefootnote{\arabic{footnote}}

\abstract{
Equilibrium and nonequilibrium states of matter can exhibit fundamentally different behavior~\cite{Odor04}. 
A key example is the Kardar--Parisi--Zhang universality class in two spatial dimensions (2D KPZ), where microscopic deviations from equilibrium give rise to macroscopic scaling laws without equilibrium counterparts~\cite{Kardar86,Krug97}. 
While extensively studied theoretically, direct experimental evidence of 2D KPZ scaling has remained limited to interface growth so far~\cite{HalpinHealy14, Takeuchi18}. 
Here, we report the observation of universal scaling consistent with the KPZ universality class in 2D exciton--polariton condensates --- quantum fluids of light that are inherently driven and dissipative, thus breaking equilibrium conditions. 
Using momentum--resolved photoluminescence spectroscopy as well as space-- and time--resolved interferometry, we probe the phase correlations across microscopically different systems, varying drive conditions in two distinct lattice geometries. 
Our analysis reveals correlation dynamics and scaling exponents in excellent agreement with 2D KPZ predictions. 
These results establish exciton--polariton condensates as a robust experimental platform for exploring 2D nonequilibrium universality quantitatively, and open new avenues for investigating the emergence of coherence in interacting quantum systems far from equilibrium.
}

\maketitle

\section*{Introduction}\label{introduction}
Universality forms the foundation of our quantitative understanding of many--particle systems. A particularly striking manifestation appears in two--dimensional systems with continuous symmetries, such as phase rotations or spin orientations.
Their phase diagram follows the Berezhinskii--Kosterlitz--Thouless (BKT) scenario~\cite{Berezinsky70, Kosterlitz73}, characterized by an algebraic decay of long--distance correlations in a low temperature critical phase.
This framework assumes thermodynamic equilibrium. 
However, recent advances in solid--state~\cite{Bloch22, Huebener21, Carusotto13} and synthetic quantum systems~\cite{Mivehvar21, Monroe21, Preskill18} actively utilise nonequilibrium conditions --- for example, through driving by laser light~\cite{Bloch22}. 
This raises a key question: How robust is the macroscopic BKT phenomenology under a breaking of equilibrium conditions at the microscopic scale? 

Here, we demonstrate that out of equilibrium, the BKT paradigm can be superseded by a distinct yet equally universal behaviour. 
Our experiments provide first evidence that 2D Kardar--Parisi--Zhang (KPZ) scaling~\cite{Kardar86} --- a cornerstone of nonequilibrium statistical mechanics~\cite{Healy97,Krug97, Takeuchi18} --- governs the phase correlations of negative mass condensates in two--dimensional exciton--polariton lattices.
Originally introduced to describe stochastic interface growth, the KPZ equation finds an unexpected counterpart in the condensate’s phase dynamics, where the phase itself plays the role of the fluctuating interface~\cite{Altman15}. 
We observe this universality across two distinct Bravais lattices.

Our result provides unambiguous macroscopic evidence of microscopic detailed balance breaking.
KPZ scaling is often associated with one--dimensional systems, such as bacterial colony growth~\cite{Wakita97} or fire fronts~\cite{Maunuksela97}. 
Quantitative results have been reported for liquid crystal turbulence~\cite{Takeuchi10, Takeuchi2011}, and its phase realization was demonstrated in exciton--polaritons~\cite{Fontaine22}. 
While these platforms constitute nonequilibrium systems, KPZ scaling is not necessarily tied to this circumstance in one dimension~\cite{Kamenev23, Tauber14a}. 
For instance, it also appears in equilibrium fluctuating hydrodynamics~\cite{Spohn14}, as seen in high--temperature integrable magnets~\cite{Scheie21, Wei22, Keenan22}. 
Since the experimental observation of KPZ scaling in a one--dimensional exciton--polariton condensate by \citet{Fontaine22}, theoretical progress has also been made in the understanding of KPZ scaling in 2D exciton--polariton condensates~\cite{Deligiannis22}. 
Experimental realizations of two--dimensional KPZ physics --- lacking an equilibrium counterpart --- have instead remained scarce, and limited to interface growth so far~\cite{Eklund91,Ojeda00, HalpinHealy14}.
No full spatiotemporal scaling collapse has been reported to date. 
Our work, demonstrating 2D KPZ universality in nonequilibrium condensate dynamics, marks not only a first step toward its quantitative characterization, but also significantly expands the range of physical platforms where it can be explored. Furthermore, it establishes two--dimensional driven--dissipative Bose condensates as genuine non--thermal phases of quantum matter.

\section*{Driven--Dissipative Condensates}

Strong light--matter coupling in semiconductor microcavities gives rise to exciton--polaritons --- hybrid quasiparticles formed by photons and excitons~\cite{Weisbuch92}. 
These bosonic states emerge when the interaction strength between cavity--confined photons and electronic excitations in the semiconductor exceeds their respective loss rates.
Typically, quantum wells are embedded within a Fabry–Pérot resonator formed by distributed Bragg reflectors (DBRs), enabling energy exchange between photons and the quantum well excitons and resulting in new part light, part matter degrees of freedom: the lower and upper polaritons.

Under sufficient excitation power, these polaritons undergo a nonequilibrium phase transition analogous to, but distinct from, Bose–Einstein condensation, marked by a macroscopic occupation of a single quantum state~\cite{Kasprzak06}. 
The transition is commonly driven via incoherent pumping --- optical or electrical --- which populates a high--energy exciton reservoir. 
Through stimulated scattering, polaritons relax toward lower energy states, and above a critical density, coherence spontaneously emerges~\cite{Bloch22}.

Unlike equilibrium condensates, where particle number is conserved and the system minimizes free energy, polariton condensates exist in a driven--dissipative flux equilibrium state, continuously exchanging energy and particles with the environment. 
This nonequilibrium nature leads to unique condensate behaviour: the system's stationary state balances losses and the pumping instead of minimizing free energy, and the condensate typically resides at a finite frequency.
The resulting phase dynamics are dominated by long--wavelength Goldstone fluctuations associated with the spontaneous breaking of a quantum mechanical U(1) symmetry.

\section*{KPZ Universality}

The interplay between drive and dissipation in this system is captured by a rate equation for the pumped reservoir occupation $n_\mathrm{R}(\mathbf{r},t)$ coupled to a Gross--Pitaevskii equation describing the lower polariton--field $\psi = \psi(\mathbf{r}, t)$ \cite{Wouters2007},
\begin{equation}
\begin{split}
&i \hbar \partial_t \psi = \left[- \frac{\hbar^2}{2 m} \nabla^2 - \frac{i \hbar}{2} \left(\gamma-\gamma_2\nabla^2\right)+ g|\psi|^2+2g_\mathrm{R} n_\mathrm{R}+ \frac{i \hbar}{2} Rn_\mathrm{R} \right]\psi + \hbar \xi(\mathbf{r}, t), \\
&\partial_t n_\mathrm{R}(\mathbf{r}, t) =P-(\gamma_\mathrm{R}+R|\psi|^2 ) \cdot n_\mathrm{R}(\mathbf{r}, t).
\end{split}\label{eq:cGPE}
\end{equation}
Finite temperature leads to white noise $\xi(\mathbf{r}, t)$, and e.g. photon leakage out of the cavity and other decay processes manifest in loss rates $\gamma$ and $\gamma_{\mathrm{R}}$. The momentum dependence of the polariton loss as well as their kinetic energy is modelled by a leading order derivative expansion around the wave vector at which the system condenses.
The momentum dependence of the polariton loss is captured to lowest order by the effective mass of the polaritons, $m$, and an effective diffusion constant $\gamma_2$. The elastic polariton scattering rate within the lower branch is $g$, and $g_\mathrm{R}$ describes the respective scattering with reservoir polaritons. The crucial incoherent, stimulated scattering from the reservoir to the lower branch polariton is given by $R$, and the laser pump of the reservoir modes with power $P$.\\

The polariton correlation function is observable through the first--order coherence function
\begin{align}
g^{(1)}(\Delta \mathbf{r}, \Delta t) = \frac{\langle\psi ^*(\mathbf{r},t_0)  \psi(-\mathbf{r},t_0+\Delta t)\rangle}{(\sqrt{|\psi (\mathbf{r},t_0)|^2})(\sqrt{|\psi (-\mathbf{r},t_0+\Delta t)|^2)}},
\label{eq:g1}
\end{align}
where $g^{(1)}(\mathbf{r}, -\mathbf{r}, \Delta t)\equiv g^{(1)}(\Delta \mathbf{r}, \Delta t)$.

Above the condensation threshold, $P>P_{\mathrm{th}}$, the many--body fluctuations of $\psi (\mathbf{r},t)= \sqrt{n(\mathbf{r},t)}e^{i\theta (\mathbf{r},t)}$ are dominated by the soft phase mode $\theta(\mathbf{r},t)$. Its dynamics can be derived by adiabatically eliminating the fast reservoir as well as the gapped density fluctuations~\cite{Altman15, Deligiannis22, Gladilin2014, Helluin24}. The resulting dynamics does not obey free diffusion, but rather is subject to the nonlinear KPZ equation,
\begin{align}
    \partial_t \theta(\mathbf{r}, t) = \nu \nabla^2 \theta(\mathbf{r}, t) + \frac{\lambda}{2} \left[\nabla \theta(\mathbf{r}, t)\right]^2 + \eta(\mathbf{r}, t) . 
\label{eq:KPZ-main}
\end{align}
The KPZ nonlinearity parametery $\lambda$, which in two spatial dimensions is strictly forbidden in equilibrium~\cite{Kamenev23, Tauber14a}, dominates the long distance behaviour of the phase fluctuations. Thus, in an infinite system, the phase dynamics does not reflect the diffusive behaviour indicated by $\nu$ and the Gaussian white noise $\langle\eta(\mathbf{r},t)\eta(\mathbf{r'},t')\rangle=2D\delta(\mathbf{r}-\mathbf{r'})\delta(t-t')$, but realizes the KPZ universality class~\cite{Deligiannis22, Fontaine22}
\begin{align}
-\log{g^{(1)}(|\Delta \mathbf{r}|,\Delta t)} = A|\Delta t|^{2\beta}\mathcal{C}\left(\frac{|\Delta \mathbf{r}|}{\Delta t^{1/z}}\right) \Rightarrow 
\begin{cases}
\log{g^{(1)}(0,\Delta t)} \sim A\Delta t^{2\beta}\\
\log{g^{(1)}(|\Delta \mathbf{r}|,0)}\sim -B|\Delta \mathbf{r}|^{2\chi}
\end{cases}
\label{eq: Scaling Function}
\end{align}
with the $2$D universal exponents $\beta\approx 0.24, \,\chi\approx 0.39$~\cite{HalpinHealy2012,Pagnani2015,Oliveira2022}, dynamical critical exponent $z=\chi/\beta$ and universal scaling function $\mathcal{C}(y)$. $A,B$ are nonuniversal constants. 

At finite length scales the KPZ scaling is observable when the dimensionless KPZ coupling  $g_{\mathrm{KPZ}}=\lambda^2 D/\nu^{3}$ is large. For smaller KPZ strengths, the phase fluctuations will be diffusive at finite scales~\cite{Claude25}. 
Above the threshold $P>P_{\mathrm{th}}$, the KPZ coupling strength decreases with lasing power $g_{\mathrm{KPZ}}\propto (P-P_{\mathrm{th}})^{-1}$ and we therefore expect to observe a KPZ scaling regime just above the lasing threshold according to recent numerical simulations~\cite{Deligiannis22, Helluin24}. 
Since the experimental system in question has a finite size, we expect the KPZ scaling to be observable, when the value of the dimensionless KPZ coupling $g_{\mathrm{KPZ}}=\lambda^2 D/\nu^{3}\propto (1-P_{\mathrm{th}}/P)^{-1}$ is large.\\
While the original KPZ equation describes an interface that grows indefinitely, the phase in \eqref{eq:KPZ-main} is a compact variable. 
This allows for the formation of spatial vortex defects in two dimensions.
At infinitely large scales, the noise in presence of the KPZ fluctuations will always lead to an unbinding of vortices which destroy coherence at very long distances~\cite{Sieberer_2016, Wachtel_2016, Zamora2017}. 

\section*{Experimental Realization}

\begin{figure}[t!]
\centering
\includegraphics[width=1\textwidth]{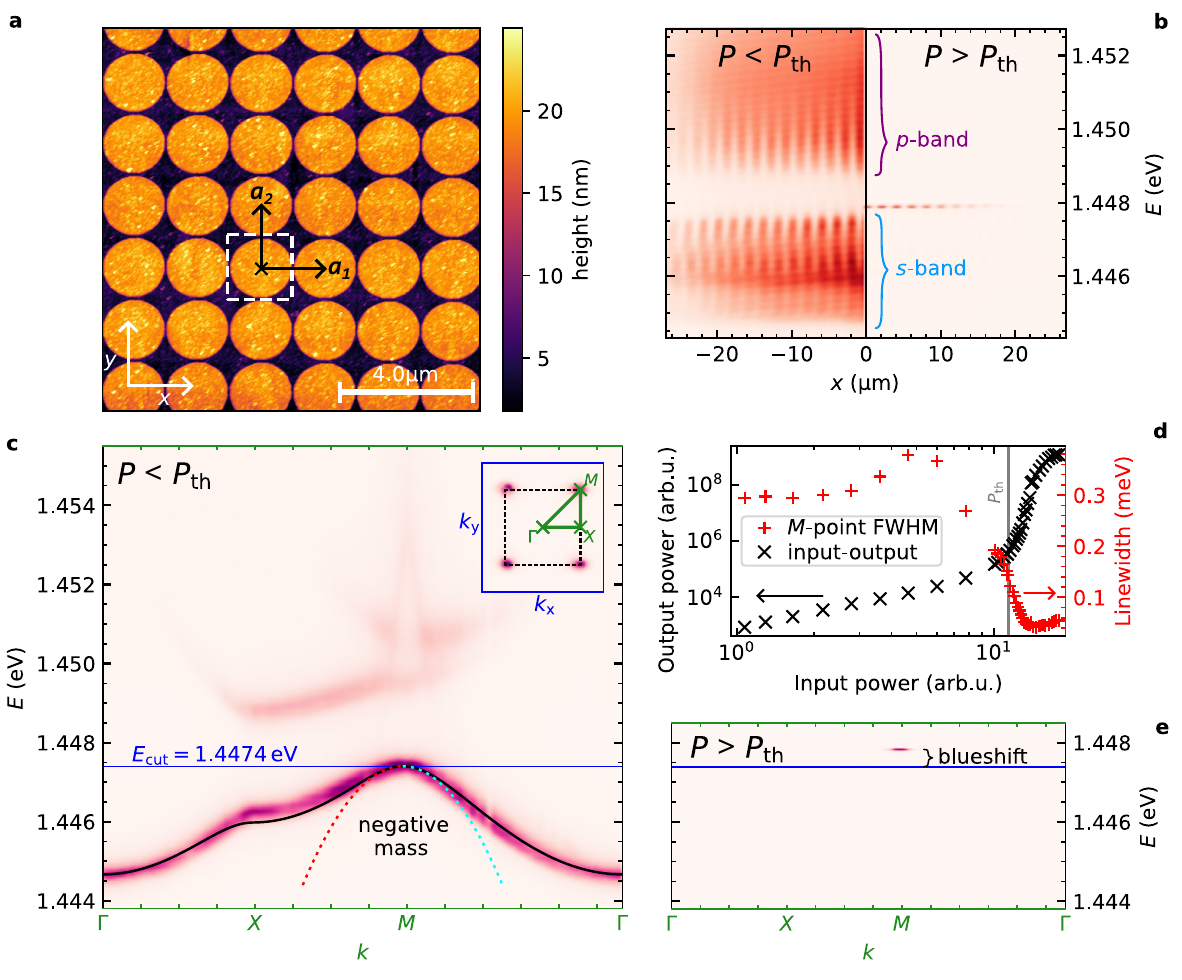}
\caption{
\textbf{Polariton square lattice characteristics. a}, Atomic force microscope (AFM) image of the microcavity patterned into a square lattice. The white box shows the unit cell spanned by the lattice vectors $\boldsymbol{a_1}$ and $\boldsymbol{a_2}$. \textbf{b}, Real--space PL spectrum of the polariton lattice showing the $s$-- and $p$--band at ($p_\mathrm{r}\equiv P/P_\mathrm{th}=0.2$) and ($p_\mathrm{r}=1.1$), indicating occupation of single energy state above $P_\mathrm{th}$. \textbf{c}, Band structure (far field) of the square lattice along a high symmetry path for $p_\mathrm{r}=0.2$. The inset highlights the $M$--points in an isoenergy cut at $E_\mathrm{cut}=1.4474\,\mathrm{eV}$. \textbf{d}, Excitation power dependent nonlinear increase of emission from the $M$--point accompanied by a reduction in linewidth, revealing polariton condensation. \textbf{e}, $M$--point condensate at $p_\mathrm{r}=1.1$ illustrating the macroscopic occupation of negative--effective--mass state.
}

\label{fig1:Band-structure}
\end{figure}

To demonstrate 2D KPZ scaling, we must accurately measure the spatiotemporal evolution of the first--order coherence function 
$g^{(1)}(\Delta \mathbf{r}, \Delta t)$ of the condensates. 
According to \eqref{eq: Scaling Function}, $g^{(1)}$ should exhibit a decay governed by the universal KPZ exponents in 2D, which is revealed by a collapse of the data onto the scaling function. 
Realizing this experimentally requires a large, spatially extended, and stable condensate to access a wide range of spatial separations with sufficient coherence~\cite{Zamora2017, Ferrier2022}.
We achieve this by tuning condensate formation at engineered negative--effective--mass states using polariton lattices~\cite{Schneider16}, where the polariton--polariton interactions are effectively tuned to be repulsive.
In contrast, simple planar microcavities with positive--mass states typically suffer from attractive polariton--polariton interactions that lead to condensate destabilization~\cite{Baboux18, Fontaine22}. 
In addition, the discretization imposed by the lattice geometry suppresses the formation of vortices~\cite{Deligiannis22}.
Here, 2D square (Figure\,\ref{fig1:Band-structure}a) and triangular (extended data Figure\,\ref{fig4: Traingular Lattice}a) lattices of micro--resonators were fabricated on a planar semiconductor microcavity sample by introducing a local thickness variation, with a monoatomic basis (sample and fabrication details in methods section).

We probe the polariton dynamics in the lattices using real-- and momentum--space resolved photoluminescence (PL) spectroscopy.
The lattices are incoherently excited by a continuous--wave laser shaped into a circular flat top profile using a spatial light modulator (see methods and supplementary information for details).

In the real--space resolved PL spectrum of the square lattice below the condensation threshold ($P_\mathrm{th}$) (Figure\,\ref{fig1:Band-structure}b), both the $s$-- and $p$--bands of the system are visible.
Above $P_\mathrm{th}$, the intensity solely originates from the top of the lowest energy $s$--band, which for a square lattice corresponds to $M$--points of the Brillouin zone, suggesting only one energy state is occupied.
The band structure of the square lattice below $P_\mathrm{th}$ is shown in Figure\,\ref{fig1:Band-structure}c. 
An isoenergy cut at $E_\mathrm{cut} = 1.4474\,\mathrm{eV}$ of the band structure is plotted in the inset, showing that this energy corresponds to the emission from the corners of the first Brillouin zone --- the $M$--points.
$P_\mathrm{th}$ is determined from the PL input--output characteristics  ($\times$, black) (Figure\,\ref{fig1:Band-structure}d), exhibiting a nonlinear increase at the $M$--point, along with a sharp linewidth reduction ($+$, red) --- hallmarks of polariton condensation.
As the excitation power surpasses $P_\mathrm{th}$, emission becomes localized exclusively at the negative--effective--mass state at the $M$--point of the Brillouin zone (Figure\,\ref{fig1:Band-structure}e), accompanied by a noticeable blueshift, indicating condensation (see extended data Figure\,\ref{fig4: Traingular Lattice} for details on the triangular lattice).
The excitonic fraction of the condensed state is $\approx 7.5\%$.

\begin{figure}[t!]
\centering
\includegraphics[width=1\textwidth]{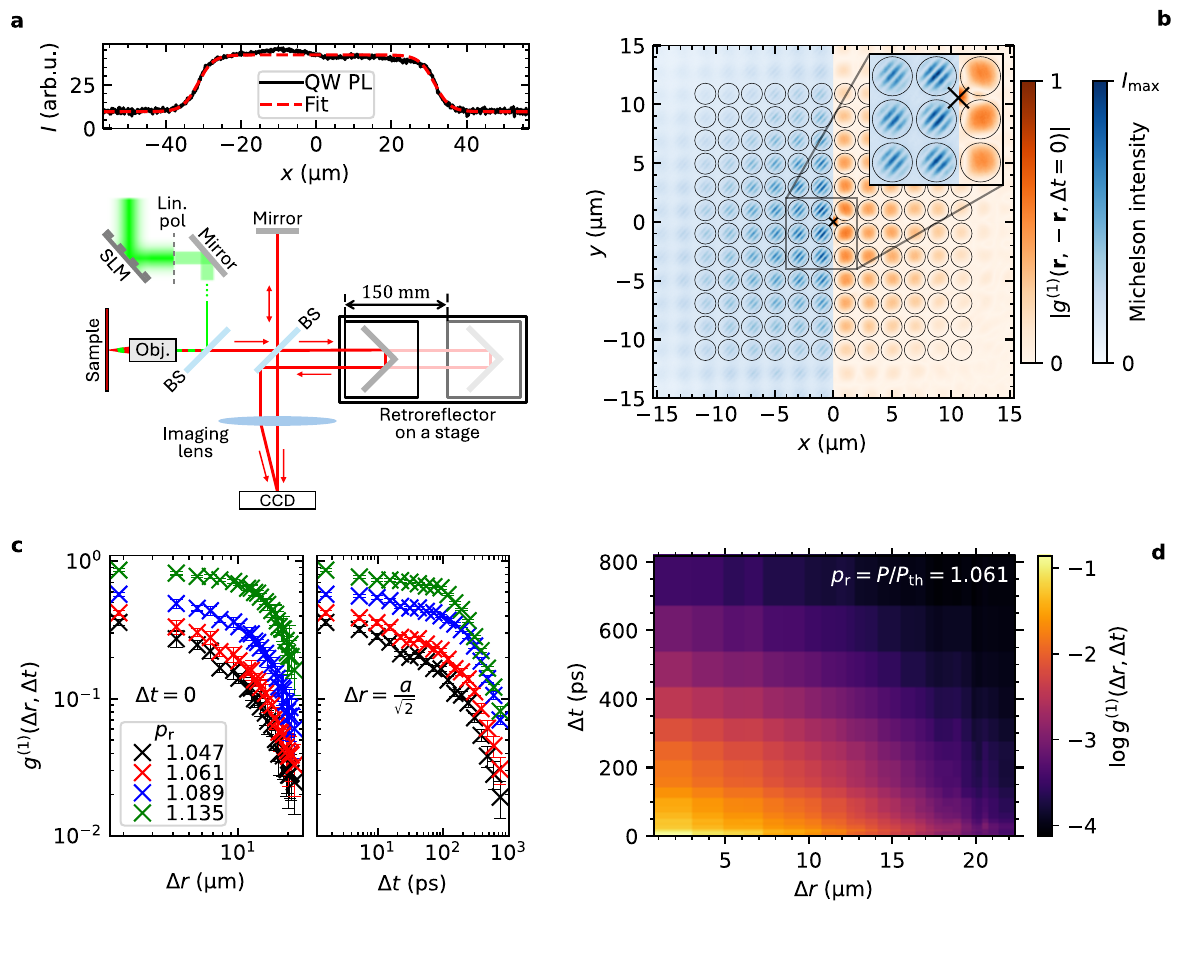}
\caption{
\textbf{Experimental access to the first order correlation function. a, Top}: Quantum well emission excited by the flat top laser spot. \textbf{Bottom}: Simplified sketch of the experimental setup.  \textbf{b}, \textbf{Left}: Interferogram at $p_\mathrm{r}=1.061$ showing fringes that exhibit a carrier wave modulation in the coherent part, at the temporal autocorrelation point $\Delta t = 0$ in the 2D square lattice $M$--point condensate. \textbf{Right}: First order correlation function $g^{(1)}(\Delta r, \Delta t = 0)$ obtained for the same interferogram. \textbf{c}, Variation of $g^{(1)}(\Delta r, \Delta t)$ at (left) ($\Delta t = 0$) and (right) ($\Delta r = a/\sqrt{2}$) with input powers slightly above $P_\mathrm{th}$. \textbf{d}, Spatiotemporal map of interpolated $g^{(1)}$ at constant output power $p_\mathrm{r}=1.061$.} 
\label{fig2: Coherence}
\end{figure}

To measure $g^{(1)}(\Delta \mathbf{r},\Delta t)$, we use a Michelson interferometer with a retroreflector (Figure\,\ref{fig2: Coherence}a and methods section and supplementary information for details).
Due to the point reflection correlation introduced by the retroreflector, we get $|\Delta \mathbf{r}| = \Delta r$.
Figure\,\ref{fig2: Coherence}b shows a representative interferogram and the extracted $g^{(1)}$ at $\Delta t = 0$ for condensate at the $M$--point of the square lattice with an excitation power of $p_\mathrm{r}\equiv P/P_\mathrm{th}= 1.061$ (see extended data Figure\,\ref{fig4: Traingular Lattice}b for triangular lattice).
The spatial autocorrelation point ($\Delta r = 0$) is set to be the high symmetry points between the pillars (centre of a pillar for the triangular lattice; marked by a cross) to preserve lattice symmetries. 
This procedure is repeated across $\Delta t$ to construct the full spatiotemporal map of $g^{(1)}(\Delta r, \Delta t)$. 

Slightly above threshold, the system exhibits a strong nonlinearity, where small variations in input power lead to large changes in output intensity (Figure\,\ref{fig1:Band-structure}d).
As a result, even minor drifts in excitation power during an interferometric scan can significantly affect the measured $g^{(1)}$ (Figure\,\ref{fig2: Coherence}c).
To mitigate this effect, we perform a power--dependent series at each $\Delta t$, measuring $g^{(1)}$ for a range of input powers and interpolate the result to a constant output intensity.
This procedure ensures that fluctuations in the measured coherence arise from the intrinsic dynamics rather than power drift. 
Further details are provided in the supplementary information.
A representative map of interpolated $g^{(1)}(\Delta r,\Delta t)$ at $p_\mathrm{r} = 1.061$ for the square lattice is shown in Figure\,\ref{fig2: Coherence}d.

\section*{Universal Scaling}

We demonstrate KPZ universality by the scaling collapse of the coherence data onto the universal scaling function obtained from a direct simulation of the KPZ equation \eqref{eq:KPZ-main} (see methods section). 
The collapse of the data onto the scaling function described by \eqref{eq: Scaling Function} provides strong evidence of KPZ universality, as it not only takes into account the scaling exponents but also the form of the scaling function.

\begin{figure}[t!]
\centering
\includegraphics[width=1\textwidth]{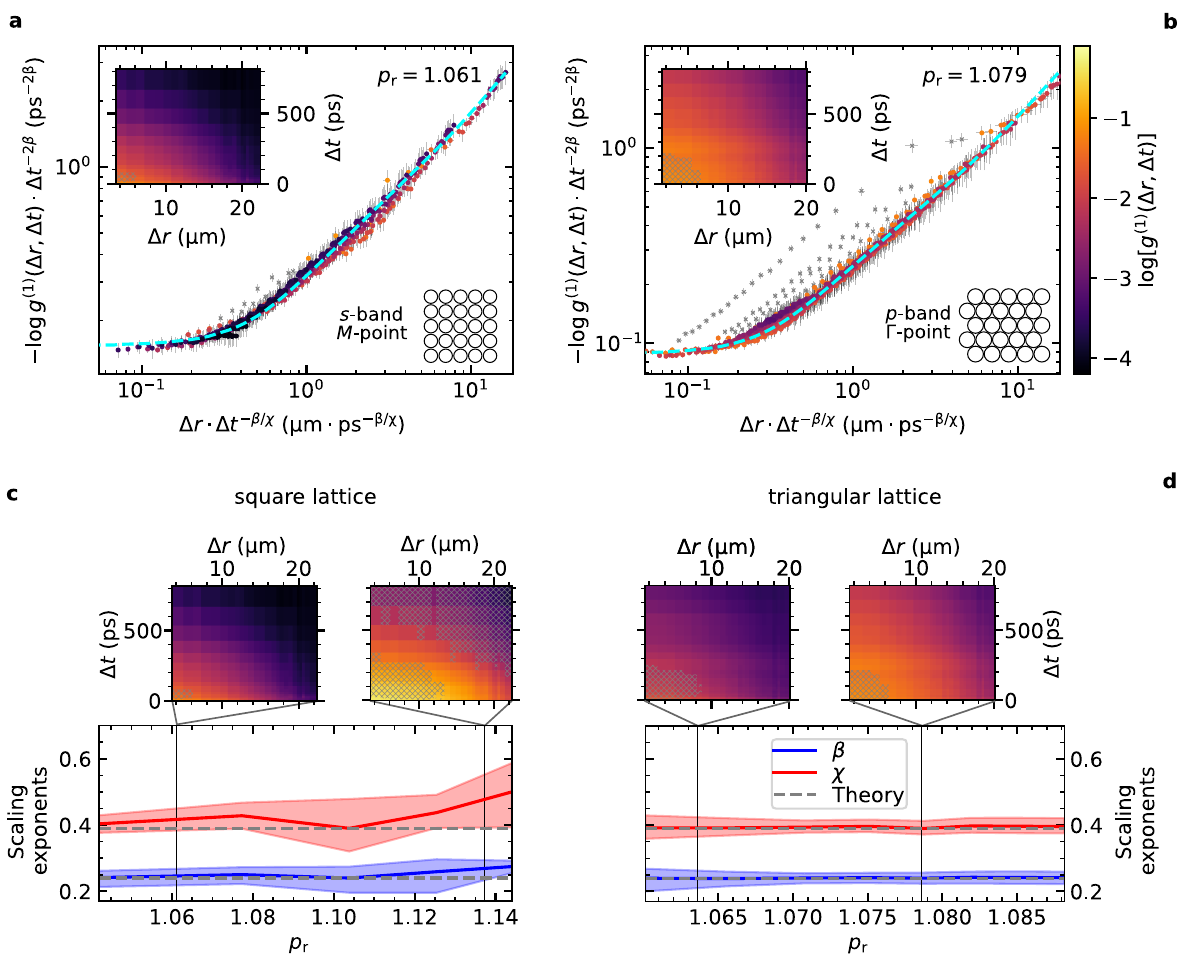}
\caption{
\textbf{2D KPZ Universal scaling. a} and \textbf{b}, Collapse of the rescaled experimental correlation data onto the numerically simulated 2D KPZ universal scaling function (cyan). The collapse of the data is a hallmark of the KPZ universal scaling behaviour. The data outside the $\pm3\sigma$ confidence window is plotted in grey ($\times$--marker). 
The inset shows the region of $g^{(1)}(\Delta r, \Delta t)$ that fits the KPZ scaling function. The data outside the confidence window is hatched grey in the inset.
\textbf{a}, Scaling collapse of the square lattice $s$--band $M$--point condensate at $p_\mathrm{r} = 1.061$.
\textbf{b}, Scaling collapse of the triangular lattice $p$--band $\Gamma$--point condensate at $p_\mathrm{r} = 1.079$.
\textbf{c} and \textbf{d}, Fitted scaling exponents $\beta$ and $\chi$. The insets on top are equivalent to \textbf{a} and \textbf{b}.  
\textbf{c}, Scaling exponents for the $s$--band $M$--point condensate of the square lattice with $p_\mathrm{r}=P/P_\mathrm{th}$.
The two insets at different excitation powers demonstrate the shrinking of the KPZ scaling region in the 2D $g^{(1)}(\Delta r, \Delta t)$ map.
\textbf{d}, Scaling exponents for the triangular lattice $p$--band $\Gamma$--point condensate. It is to be noted that the accessible power range for the triangular lattice is smaller than for the square lattice.
}
\label{fig3: KPZ Scaling}
\end{figure}

\begin{figure}[H]
\centering
\includegraphics[width=1\textwidth]{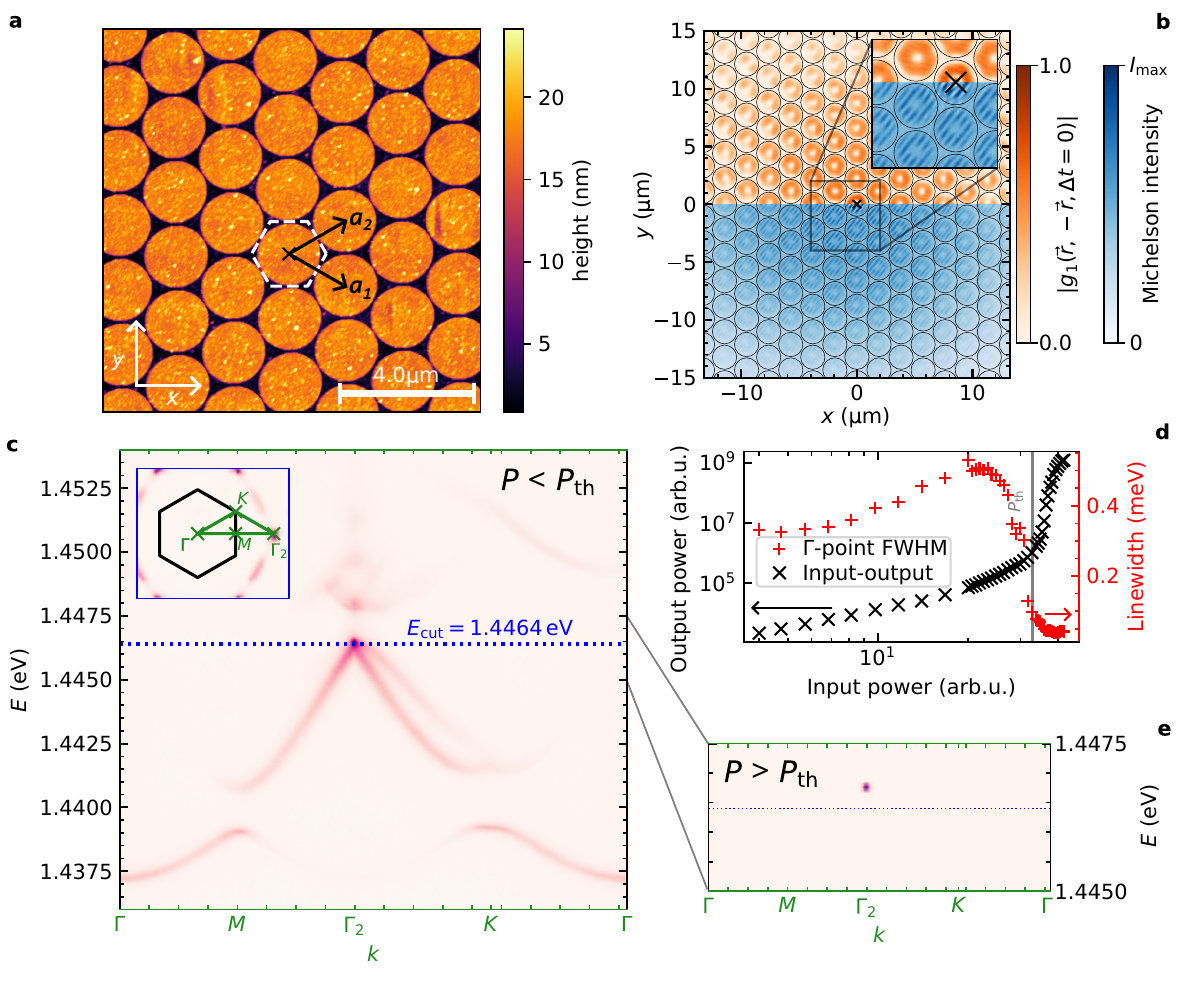}
\caption{
\textbf{Triangular lattice geometry, band structure and interferometry. a}, AFM micrograph of the triangular lattice. The white hexagon shows the unit cell with lattice vectors $\boldsymbol{a_1}$ and $\boldsymbol{a_2}$. \textbf{b}, (bottom) shows an exemplary interferogram at $p_\mathrm{r} =1.1$ and $\Delta t= 0$. The centre of the pillars shows no emission, reflecting the radial intensity distribution of the $p$--orbitals. (Top) shows the extracted $g^{(1)}(\Delta r, \Delta t=0)$. The autocorrelation point is marked as $\boldsymbol{\times}$ and is chosen so that all the symmetries of the lattice are preserved. \textbf{c}, band structure of the triangular lattice probed by Fourier PL below the condensation threshold. The inset shows the isoenergy cut at $E_\mathrm{cut} = 1.4464\,\mathrm{eV}$ demonstrating the emission from the $\Gamma$--points of higher order Brillouin zones. \textbf{d}, Input--output characteristics of the emission from the $\Gamma$--points of the $p$--band showing hallmarks of polariton condensation above $P_{\mathrm{th}}$. \textbf{e}, $k$--space PL at $p_\mathrm{r}= 1.1$ showing localized emission and blueshift indicating polariton condensation. The blue dashed line has the same energy as in \textbf{c}, $E_\mathrm{cut} = 1.4464\,\mathrm{eV}$. The condensed state has an excitonic fraction of $\approx 5.9\%$.
}
\label{fig4: Traingular Lattice}
\end{figure}

\begin{figure}[H]
\centering
\includegraphics[width=1\textwidth]{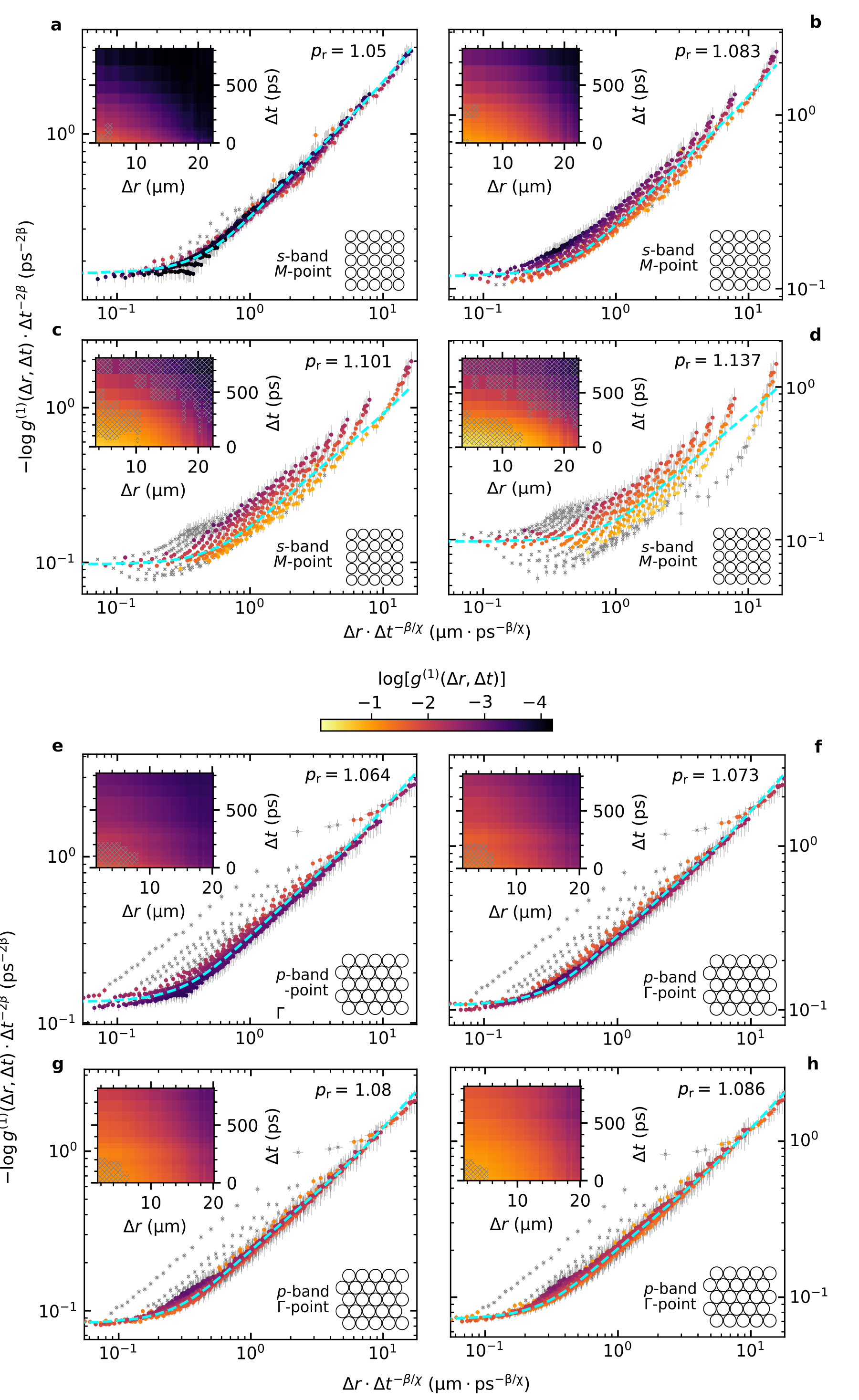}
\caption{
\textbf{Scaling collapse for different input powers. a-d}, Scaling collapse at different powers for the square lattice $M$--point condensate. For increasing power, an increasing deviation from the scaling function is observed. The hatched region denoting the points lying outside the confidence window of $3\sigma$ also increase indicating deviation from KPZ scaling. \textbf{e-h}, Scaling collapse at different powers for the $\Gamma$--point condensate in the $p$--band of the triangular lattice. In the range of the accessible power, the data follows the KPZ scaling strikingly well.
}
\label{fig5: scaling collapse}
\end{figure}

Therefore, we plot the rescaled coherence function $-\log\left(g^{(1)}(\Delta r, \Delta t)\right) \cdot \Delta t^{-2\beta}$ against $\Delta r \cdot \Delta t^{-\beta/\chi}$
following the scaling form in \eqref{eq: Scaling Function}. 
The two non--universal prefactors, $A$ and $B$, allow for vertical and horizontal shifts of the scaling function and are fitted using orthogonal distance regression (ODR) across the full dataset in $\Delta r$ and $\Delta t$.
Figures\,\ref{fig3: KPZ Scaling}a and b display the scaling collapse of the experimental data onto the theoretical KPZ curve for the square--lattice $M$--point condensate at $p_\mathrm{r} = 1.061$ and for the triangular--lattice $\Gamma$--point condensate of the $p$--band at $p_\mathrm{r} = 1.079$, respectively. 
Points deviating by more than three standard deviations ($3\sigma$ in the ODR) from the fitted curve are shown in grey.
In both cases, the majority of the data fall within the $3\sigma$ range, indicating statistically robust agreement with the predicted KPZ scaling function. 
The locations of the excluded points in the $(\Delta r, \Delta t)$--plane are shown as hatched regions in the insets. These deviations are confined to small $\Delta r$ and $\Delta t$, possibly reflecting a suppression of KPZ phase fluctuations by an incoherent background from the uncondensed reservoir~\cite{Fontaine22}.
The scaling collapse of the experimental data with different excitation powers is shown in extended data Figure\,\ref{fig5: scaling collapse}a-d for the square lattice and c-f for the triangular lattice.
Specifically, the deviation of the square lattice correlation from the KPZ regime becomes evident at higher excitation powers (Figure\,\ref{fig5: scaling collapse}c,d).

We now extract the experimental values of the KPZ scaling exponents $\chi$ and $\beta$ by restricting the ODR fit to the KPZ regions (non--hatched areas in the insets of extended data Figure\,\ref{fig5: scaling collapse}a-h), as identified by the procedure described above.
Within this region, the rescaled $ g^{(1)}(\Delta r, \Delta t) $ data is fitted to the universal KPZ scaling function with $\chi$ and $\beta$ treated as free parameters.
Figure\,\ref{fig3: KPZ Scaling}c shows the dependence of the extracted exponents on excitation power for the square lattice. 
At $p_\mathrm{r} = 1.061$, the extracted scaling exponents are $\chi=0.417\pm0.034$ and $\beta=0.246\pm0.028$, close to the theoretically predicted values of $0.39$ and $0.24$~\cite{Deligiannis22}, respectively.
At higher powers, the number of data points excluded from the KPZ region (hatched areas) increases, indicating a departure from the KPZ prediction.
This deviation can be attributed to the decrease of the KPZ coupling $g_\mathrm{KPZ}$ as the excitation power increases.
The extracted exponents for the triangular lattice are presented in Figure\,\ref{fig3: KPZ Scaling}d.
For a power of $p_\mathrm{r}=1.064$, $\chi = 0.392\pm0.032$ and $\beta = 0.239\pm0.029$.
Over the experimentally accessible power range, the triangular lattice results show excellent agreement with KPZ theory.
As illustrated in the corresponding insets, the extent of the KPZ region in the $(\Delta r, \Delta t)$--plane remains largely unchanged.

These results confirm that, despite differing microscopic parameters, the phase dynamics of polariton condensates in both lattice systems exhibit the same 2D universal KPZ scaling. The observation of 2D KPZ scaling underscores the inherently nonequilibrium, driven--dissipative nature of polariton condensates, setting them apart from implementations of 1D KPZ scaling, where equilibrium realizations exist. 
\\
\section*{Conclusion and Outlook}

We have demonstrated the emergence of KPZ universal scaling in 2D in out--of--equilibrium, driven--dissipative polariton condensates. By probing the coherence properties of two microscopically distinct 2D polariton lattices, we access both spatial and temporal correlations. This comprehensive experimental control allows us, for the first time, to construct the universal KPZ scaling function directly from measured data, in excellent agreement with theoretical predictions.

Several compelling directions are now within reach. While our current investigations focus on the regime near the condensation threshold --- where KPZ physics governs long--distance correlations --- exploring deeper into the condensed phase or at a higher noise level will allow us to resolve the dynamics of vortex defects. 
These topological excitations are expected to interplay with both nonequilibrium KPZ and equilibrium BKT physics~\cite{Sieberer_2016,Wachtel_2016, Deligiannis22, Helluin24}, offering a path to chart the full phase diagram and scaling regimes of two--dimensional driven open quantum systems. 
In addition, the tunability of our lattice platform offers the opportunity to engineer anisotropies that may induce a transition between a nonequilibrium KPZ phase and an emergent equilibrium--like phase, as theoretically predicted for systems with strong anisotropy~\cite{Altman15}. The universal properties of such an equilibrium--to--nonequilibrium transition remain largely unexplored.

By combining advanced spectroscopic tools with cutting--edge sample engineering, exciton--polariton condensates in two dimensions are poised to become powerful `nonequilibrium simulators' for controlled, quantitative exploration of far--from--equilibrium universality, bridging the domains of optics and many--body physics.

\newpage
\section*{Methods}\label{methods}
\subsection*{Sample growth}
The sample was fabricated using an etch and overgrowth protocol in a Molecular Beam Epitaxy (MBE) system to introduce local cavity thickness variations~\cite{Daif06, Winkler15}. 
Growth began with a bottom DBR comprising $44.5$ pairs of $\mathrm{AlAs}/\mathrm{GaAs}$ layers of $\lambda/4$ thickness. A $\lambda$-thick $\mathrm{GaAs}$ cavity containing three $\mathrm{In_{0.06}Ga_{0.94}As}$ quantum wells, positioned at the antinodes of the optical field, forms the active region.
Following this, the sample was removed from vacuum and spin--coated with a positive polymethyl methacrylate (PMMA) resist. 
The required lattice pattern was written via electron--beam lithography. 
After development, a $20\,\mathrm{nm}$ aluminium film was thermally evaporated to form a etch mask. 
The remaining resist was lifted off using pyrrolidone, and the exposed sample was wet--etched in an $\mathrm{H_2O:H_2O_2}(30\%):\mathrm{H_2SO_4}(96\%)=(800:4:1)$ solution with etch depth controlled by time. 
The aluminium mask was subsequently removed using a $1\% \,\mathrm{NaOH}$ solution, and the surface was cleaned by immersion in $96\%\, \mathrm{H_2SO_4}$ for two minutes.
The sample was then reintroduced into the MBE chamber for overgrowth.
The microcavity was completed by deposition of a top DBR consisting of 40 pairs of $\mathrm{AlAs/Al_{0.2}Ga_{0.8}As}$ layers of quarter--wavelength thickness.

The etching process locally reduces the cavity thickness, resulting in a blue--shifted cavity mode and thus creating lateral photonic confinement in the patterned regions. 
Each lattice site functions as an optical resonator, and their periodic coupling gives rise to hybridized, delocalized states and the formation of polaritonic band structures. The energy separation between orbital modes yields well--defined $s$--, $p$-- and $d$-- like bands.
In our study we implemented 2D square and triangular lattices. 
For the square lattice a resonator diameter of $1.8\,\mathrm{\mu m}$ and a resonator centre--to--centre distance (also the lattice constant $a$) of $1.98\,\mu\mathrm{m}$ was implemented. 
While for the triangular lattice, the resonator diameter was $2.1\,\mathrm{\mu m}$ and the centre--to--centre distance of $2.31\,\mu\mathrm{m}$.

\subsection*{Spectroscopy and Interferometry}
A wavelength--tunable continuous--wave laser (M--squared SolsTis) was used to off--resonantly excite the microcavity in reflection geometry. The laser beam was shaped to a circular flat--top using an SLM (see supplementary information). 
Both excitation and detection were performed through the same microscope objective. For momentum--resolved (far--field, $k$--space) measurements, a $20\times$ objective with a numerical aperture $\mathrm{NA}=0.40$ was employed, while real--space imaging utilized a $50\times$ objective ($\mathrm{NA}=0.65$).

Laser incoupling was achieved via a polarizing beam splitter to maximize the optical power delivered to the sample surface. 
The sample was mounted in vacuum within a continuous--flow liquid helium cooled cryostat (Janis ST--500) and maintained at a stabilized temperature of $\approx6\,\mathrm{K}$ using a PID--controlled heater, ensuring thermal stability during extended measurement sequences.

The photoluminescence was isolated from the excitation beam using a long--pass filter.
Spectrally resolved measurements were carried out using a Czerny–Turner spectrometer (Andor Shamrock 750) coupled to a Peltier cooled CCD camera (Andor iKon--M).
Mode tomography and hyperspectral imaging were performed by mechanically scanning the image plane across the entrance slit of the spectrometer.
In hyperspectral imaging (Figure\,\ref{fig1:Band-structure}c), the back focal plane of the objective was projected onto the detector, yielding the angle-- and energy--resolved far--field emission. 
In contrast, an energy--resolved mode tomography (Figure\,\ref{fig1:Band-structure}b) directly mapped the sample surface.

For the interferometry, the real--space interferogram was projected into a high resolution CCD camera (Andor Clara). 
For the Michelson interferometry, PL emission from the sample is split up by a beam splitter and then superimposed after being reflected by a mirror and a retroreflector, respectively.
In this configuration, the retroreflector allows for a superposition of the real space PL and its point reflection, thereby correlating the points $\mathbf{r}$ and $-\mathbf{r}$, meaning $\Delta r = |\Delta \mathbf{r}|$.
Temporal correlations are obtained by adjusting the position of the retroreflector using a motorized stage, imposing a temporal delay $\Delta t$ between the two interferometer arms.
Besides inverting the image, the retroreflector also provides a parallel offset of one partial beam, that results in an angular offset after the imaging lens.
This way, the coherent part of the image is imposed with a carrier wave, resulting in interference fringes as depicted in Figure\,\ref{fig2: Coherence}b.
The wave allows for an algorithmic separation of the correlation function from the remaining incoherent signal, using a fast Fourier transformation (see supplementary information).
The shot noise induced uncertainty in $g^{(1)}$, is quantified through Monte Carlo simulations, providing a robust estimate of statistical errors (see supplementary information).

\subsection*{Theoretical simulation of KPZ equation}

Considering the KPZ equation \,\eqref{eq:KPZ-main}, the connected two--point correlation function and the universal scaling function
\begin{align}
C(\mathbf{r}-\mathbf{r}',t-t')=\langle \left(\theta(\mathbf{r},t)-\theta(\mathbf{r}',t')\right)^2\rangle_c,
\label{eq:two-point C}
\end{align}
takes the following universal scaling form in the ``rough" strong-coupling KPZ phase:
\begin{align}
C(|\Delta \mathbf{r}|,\Delta t) = A|\Delta t|^{2\beta}\mathcal{C}\left(\frac{|\Delta \mathbf{r}|}{\Delta t^{1/z}}\right) \Rightarrow 
\begin{cases}
C(0,\Delta t) \sim A\Delta t^{2\beta}\\
C(\Delta\mathbf{r},0) \sim -B|\Delta\mathbf{r}|^{2\chi}
\end{cases}
\label{eq: Scaling Function in C}
\end{align}
where we have used $|\Delta \mathbf{r}| = \mathbf{r}-\mathbf{r}'$ and $\Delta t = t-t'$, $\beta$ and $\chi$ are the universal scaling exponents associated with the KPZ universality class, and $A$ and $B$ are non--universal prefactors. In two dimensions, $\beta\approx 0.24$ and $\chi\approx 0.39$ in $d=2$~\cite{HalpinHealy2012,Pagnani2015,Oliveira2022,Deligiannis22}. Because of the presence of a Galilean symmetry in the KPZ equation, there is an exact relation between these exponents, $\beta = \chi/(2-\chi)$~\cite{Tauber14a}, and therefore only one independent exponent. 
In two dimensions, the universal scaling function has been obtained from a functional renormalization group (FRG) approach~\cite{Kloss2012} and from numerical simulation of the dynamics of time--periodic phases~\cite{Deligiannis22, Daviet2024a}.

\subsection*{Numerical simulations}
Here, we extract the universal KPZ scaling in two dimensions from a direct simulation of the KPZ equation ~\eqref{eq:KPZ-main}. In our simulations, we use a coupling of $\lambda = 3$ on a square lattice of size $L \times L$, lattice spacing $a=1$, and periodic boundary conditions. We use the Euler--Maruyama scheme in real space with a time step $dt=10^{-2}$. We average over $N=400$ realizations for a lattice length $L=1536$.

We reproduce the best practice literature values for the roughness exponent $2\chi=0.774$ \cite{Pagnani2015,Oliveira2022} using the finite size scaling of the roughness function $W(t)=\int_x\langle(\phi(x,t)-\langle\phi(x,t)\rangle)^2\rangle_c\propto L^{2\chi}$ for $t\gg L^z$  to with $2\chi= 0.775(5)$ in our simulations. The exponents can also be extracted via the asymptotes of the two point correlation function \eqref{eq: Scaling Function in C} or 
\begin{align}
    2\tilde\beta(\Delta t)=\frac{\Delta t}{C(0,\Delta t)}\frac{d}{d\Delta t} C(0,\Delta t), \quad \text{and} \quad 2\tilde\chi(\Delta \mathbf{r})=\frac{|\Delta \mathbf{r}|}{C(|\Delta \mathbf{r}|,0)}\frac{d}{d|\Delta \mathbf{r}|} C(|\Delta \mathbf{r}|,0),
\end{align}
which should reach plateaus, $\tilde \beta(\Delta t) = \beta$ and $\tilde \chi(\Delta t) = \chi$ before finite size effects set in. 
In our simulations, this yields $2\beta= 0.446(5)$, $2\chi=0.730(5)$. The finite size effects are known to be strong for the KPZ equation ~\cite{Pagnani2015}, which explains the deviation of $\chi$ from the literature value. We find that the scaling hypothesis Eq.\,\eqref{eq: Scaling Function} holds clearly for $ 15 \lesssim |\boldsymbol{x}| \lesssim 50$, and $t \in [0,1500]$. Using the exponents from the asymptotes we achieve the best scaling collapes in this regime and extract the universal scaling function $\mathcal{C}(y)$ plotted as theory curve in Figure\,\ref{fig3: KPZ Scaling}a,b in cyan.

\clearpage
\section*{Extended Data}
\subsection*{Triangular lattice}

\subsection*{Power dependent scaling collapse}

\section*{Data availability}
All datasets generated and analysed during this study are available upon reasonable
request from the corresponding authors.

\section*{Acknowledgements}
The Würzburg group acknowledges financial support by the German Research Foundation (DFG) under Germany’s Excellence Strategy–EXC2147 “ct.qmat” (project id 390858490).
The Köln group acknowledges support by the Deutsche Forschungsgemeinschaft (DFG, German Research Foundation) under Germany’s Excellence Strategy Cluster of Excellence Matter and Light for Quantum Computing (ML4Q) EXC 2004/1 390534769
and CRC 1238 project C04 number 277146847.

\section*{Author contributions}
S.W., S.Dam, C.G.M. and J.D. built the experimental setup, performed the experiments and analysed the data. S.Dam grew the samples by molecular beam epitaxy. M.E., M.K., S.W., D.L. and S.B. realized the layout, etching and nanofabrication of the samples. R.D., C.Z. and S.Diehl realized the theoretical analysis and numerical simulations. All authors participated in the scientific discussions about all aspects of the work. S.Dam and S.W. wrote the original draft of the paper with input from all authors. S.K. and S.H. conceived the idea and supervised the work.

\bibliography{sn-bibliography}

\end{document}